\documentclass[nonacm,sigconf]{acmart}

\AtBeginDocument{%
  \providecommand\BibTeX{{%
    \normalfont B\kern-0.5em{\scshape i\kern-0.25em b}\kern-0.8em\TeX}}}

\usepackage{subfigure}
\usepackage{xcolor}
\usepackage{hyperref}
\usepackage{pifont}

\newcommand{\cmark}{\ding{51}}
\newcommand{\xmark}{\ding{55}}

\begin{document}

\title{Beyond Benchmarks: Evaluating Embedding Model Similarity for Retrieval Augmented Generation Systems}

\author{Laura Caspari}
\email{laura.caspari@uni-passau.de}
\orcid{0009-0002-6670-3211}
\affiliation{%
  \institution{University of Passau}
  \city{Passau}
  \country{Germany}
}

\author{Kanishka Ghosh Dastidar}
\email{kanishka.ghoshdastidar@uni-passau.de}
\orcid{0000-0003-4171-0597}
\affiliation{%
  \institution{University of Passau}
  \city{Passau}
  \country{Germany}
}

\author{Saber Zerhoudi}
\email{saber.zerhoudi@uni-passau.de}
\orcid{0000-0003-2259-0462}
\affiliation{%
  \institution{University of Passau}
  \city{Passau}
  \country{Germany}
}

\author{Jelena Mitrovic}
\email{jelena.mitrovic@uni-passau.de}
\orcid{0000-0003-3220-8749}
\affiliation{%
  \institution{University of Passau}
  \city{Passau}
  \country{Germany}
}

\author{Michael Granitzer}
\email{michael.granitzer@uni-passau.de}
\orcid{0000-0003-3566-5507}
\affiliation{%
  \institution{University of Passau}
  \city{Passau}
  \country{Germany}
}

\renewcommand{\shortauthors}{Caspari et al.}

\begin{abstract}
   The choice of embedding model is a crucial step in the design of Retrieval Augmented Generation (RAG) systems. Given the sheer volume of available options, identifying clusters of similar models streamlines this model selection process. Relying solely on benchmark performance scores only allows for a weak assessment of model similarity. Thus, in this study, we evaluate the similarity of embedding models within the context of RAG systems. Our assessment is two-fold: We use Centered Kernel Alignment to compare embeddings on a pair-wise level. Additionally, as it is especially pertinent to RAG systems, we evaluate the similarity of retrieval results between these models using Jaccard and rank similarity. We compare different families of embedding models, including proprietary ones, across five datasets from the popular Benchmark Information Retrieval (BEIR). Through our experiments we identify clusters of models corresponding to model families, but interestingly, also some inter-family clusters. Furthermore, our analysis of top-$k$ retrieval similarity reveals high-variance at low $k$ values. We also identify possible open-source alternatives to proprietary models, with Mistral exhibiting the highest similarity to OpenAI models. 
\end{abstract}



\maketitle

\section{Motivation}

Retrieval-Augmented Generation (RAG) is an emerging paradigm that helps mitigate the problems of factual hallucination \cite{hallucination} and outdated training data \cite{outdated} of large language models (LLMs) by providing these models with access to an external, non-parametric knowledge source (e.g. a document corpus). Central to the functioning of RAG frameworks is the retrieval step, wherein a small subset of candidate documents is retrieved from the document corpus, specific to the input query or prompt. This retrieval process, known as dense-retrieval, hinges on text embeddings. Typically, the generation of these embeddings is assigned to an LLM, for which there are several options due to the rapid evolution of the field. Consequently, selecting the most suitable embedding model from an array of available choices emerges as a critical aspect in the development of RAG systems. The information to guide this choice is currently primarily limited to architectural details (which are also on occasion scarce due to the prevalence of closed models) and performance benchmarks such as the Massive Text Embedding Benchmark (MTEB) \cite{mteb}.

We posit that an analysis of the similarity of the embeddings generated by these models would significantly aid this model selection process. Given the large number of candidates and ever increasing scale of the models, a from-scratch empirical evaluation of the embedding quality of these LLMs on a particular task can incur significant costs. This challenge becomes especially pronounced when dealing with large-scale corpora comprising potentially millions of documents. While the relative performance scores of these models on benchmark datasets offer the simplified perspective of comparing a single scalar value on an array of downstream tasks, such a view of model similarity might overlook the nuances of the relative behaviour of the models \cite{similarity}. As an example, the absolute difference in precision@k between two retrieval systems only provides a  weak indication of the overlap of retrieved results. We argue that identifying clusters of models with similar behaviour would allow practitioners to construct smaller, yet diverse candidate pools of models to evaluate. Beyond model selection, as highlighted by Klabunde et al., \cite{similarity_llm}, such an analysis also facilitates  the identification of common factors contributing to strong performance, easier model ensembling, and detection of potential instances of unauthorized model reuse.

In this paper, we analyze different LLMs in terms of the similarities of the embeddings they generate. Our similarity analysis serves as an unsupervised evaluation framework for these embedding models, in contrast to performance benchmarks that require labelled data. We do this from a dual perspective - we directly compare the embeddings using representational similarity measures. Additionally, we evaluate model similarity specifically in terms of their functional impact on RAG systems i.e. we look at how similar the retrieved results are.
Our evaluation focuses on several prominent model families, to analyze similarities both within and across them. We also compare proprietary models (such as those by OpenAI or Cohere) to open-sourced ones in order to identify the most similar alternatives. Our experiments are carried out on five popular benchmark datasets to determine if similarities between models are influenced by the choice of data. Our code is available at \url{https://github.com/casparil/embedding-model-similarity}.

\section{Related Work}

Studies evaluating similarities of neural networks fall into two main categories: the first involves comparing activations of different models generated at any pair of layers for a specific input (representational similarity), while the second compares the model outputs (functional similarity). Raghu et al. \cite{svcca} and Morcos et al. \cite{pwcca} propose measures building on Canonical Correlation Analysis (CCA) \cite{cca}, a statistical technique used to find the linear relationship between two sets of variables by maximizing their correlation. Such comparisons using CCA or variants thereof can be found in several works \cite{cca_app1}, \cite{cca_app2}, \cite{cca_app3}. Beyond CCA-based measures, other works have also explored computing correlations \cite{correlation} and the mutual information \cite{mutual_info} between neurons across networks. Kornblith et al. \cite{cka} propose Centered Kernel Alignment (CKA), which they show improves over several similarity measures in identifying corresponding layers of identical networks with different initializations. A diverse range of functional similarity evaluations have also been explored in the literature. A few examples include model-stitching \cite{stitching_revisited}, \cite{model_stitching}, \cite{model_stitching_robust}, disagreement measures between output classes \cite{churn}, \cite{diffchaser}, and quantifying the similarity between the class-wise output probabilities \cite{model_diff}. We would point the reader to the survey by Klabunde et al. \cite{similarity} for a detailed overview of representational and functional similarity measures.

Recently, a few works have also focused on specifically evaluating the similarity of LLMs. While Wu et al. \cite{contextsimilarity} evaluate language models along several perspectives, such as their representational and neuron-level similarities, their evaluation pre-dates the introduction of the recent wave of large scale models. Freestone and Santu \cite{wordembedding} consider similarities of word embeddings, and evaluate if LLMs differ significantly to classical encoding models in terms of their representations. The works by Klabunde et al. \cite{similarity_llm} and Brown et al. \cite{dissimilarity} are more recent, and evaluate the representational similarity of LLMs, with the latter also considering the similarities between models of different sizes in the same model family.

Much of the literature on evaluation of LLM embeddings focuses on their performance on downstream tasks, with benchmarks such as BEIR \cite{beir} (for retrieval specifically) and MTEB \cite{mteb} providing a unified view of embedding quality across metrics and datasets. The metrics used here mostly include typical information retrieval metrics such as precision, recall, and mean reciprocal rank at certain cutoffs. Some works specifically evaluate the retrieval components in a RAG context, where they either use a dataset outside of those included in the benchmarks \cite{rageval} or where the evaluation encompasses other aspects of the retriever beyond the embedding model being used \cite{rageval2}. Another approach, that does not rely on ground-truth labels, is given by the Retrieval Augmented Generation Assessment (RAGAS) framework, which uses an LLM to determine the ratio of sentences in the retrieved context that are relevant to the answer being generated \cite{ragas}. To the best of our knowledge, there are no works that evaluate the similarity of embedding models from a retrieval perspective.

\begin{table}[]
    \centering
    \caption{The datasets used for generating embeddings with their number of queries and corpus size.}
    \begin{tabular}{lcc}
        \toprule
        Dataset Name & Queries & Corpus \\
        \midrule
        TREC-COVID & 50 & 171k \\
        NFCorpus & 323 & 3.6k \\
        FiQA-2018 & 648 & 57k \\
        ArguAna & 1406 & 8.67k \\
        SciFact & 300 & 5k \\
        \bottomrule
    \end{tabular}
    \label{tab:datasets}
\end{table}

\section{Methods}

We evaluate embedding model similarity using two approaches. The first directly compares the embeddings of text chunks generated by the models. The second approach is specific to the RAG context, where we evaluate the similarity of retrieved results for a given query. These approaches are discussed in detail in the following sections.

\begin{table*}[]
    \centering
    \caption{We compare a diverse set of open source models from different families as well as proprietary models with varying performance on MTEB.}
    \begin{tabular}{lcccc}
        \toprule
         Model & Embedding dimension & Max. Tokens & MTEB Average & Open Source \\
         \midrule
         SFR-Embedding-Mistral & 4096 & 32768 & 67.56 & \cmark \\
         mxbai-embed-large-v1 & 1024 & 512 & 64.68 & \cmark \\
         UAE-Large-V1 & 1024 & 512 & 64.64 & \cmark \\
         text-embedding-3-large & 3072 & 8191 & 64.59 & \xmark \\
         Cohere embed-english-v3.0 & 1024 & 512 & 64.47 & \xmark \\
         bge-large-en-v1.5 & 1024 & 512 & 64.23 & \cmark \\
         bge-base-en-v1.5  & 768 & 512 & 63.55 & \cmark \\
         gte-large & 1024 & 512 & 63.13 & \cmark \\
         gte-base & 768 & 512 & 62.39 & \cmark \\
         text-embedding-3-small & 1536 & 8191 & 62.26 & \xmark \\
         e5-large-v2 & 1024 & 512 & 62.25 & \cmark \\
         bge-small-en-v1.5 & 384 & 512 & 62.17 & \cmark \\
         e5-base-v2 & 768 & 512 & 61.5 & \cmark \\
         gte-small & 384 & 512 & 61.36 & \cmark \\
         e5-small-v2 & 384 & 512 & 59.93 & \cmark \\
         gtr-t5-large & 768 & 512 & 58.28 & \cmark \\
         sentence-t5-large & 768 & 512 & 57.06 & \cmark \\
         gtr-t5-base & 768 & 512 & 56.19 & \cmark \\
         sentence-t5-base & 768 & 512 & 55.27 & \cmark \\
         \bottomrule
    \end{tabular}
    \label{tab:models}
\end{table*}

\subsection{Pair-wise Embedding Similarity}

There are several metrics defined in the literature that measure representational similarity \cite{similarity}. Many of these metrics require the representation spaces of the embeddings to be compared to be aligned and/or the dimensionality of the embeddings across the models to be identical. To avoid these constraints, we pick Centered Kernel Alignment (CKA) \cite{cka} with a linear kernel as our similarity measure.

The measure computes similarity between two sets of embeddings in two steps. First, for a set of embeddings, the pair-wise similarity scores between all entries within this set are computed using the kernel function. Thus, row \textit{k} of the resulting similarity matrix contains entries representing the similarity between embedding \textit{k} and all other embeddings, including itself. Computing two such embedding similarity matrices for different models with the same number of embeddings then leads to two matrices E and E' of matching dimensions. These are compared directly in the second step with the Hilbert-Schmidt Independence Criterion (HSIC) \cite{hsic} using the following formula:

\begin{equation}
    CKA(E, E') = \frac{HSIC(E, E')}{\sqrt{HSIC(E, E)HSIC(E', E')}}
\end{equation}

The resulting similarity scores are bounded in the interval [0, 1] with a score of 1 indicating equivalent representations. CKA assumes that representations are mean-centered.

\subsection{Retrieval Similarity}
While a pair-wise comparison of embeddings offers insights into the similarities of the representations learned by these models, it does not suffice to quantify the similarities in outcomes when these embedding models are deployed for specific tasks. Therefore, in context of RAG systems, we consider the similarity of retrieved text chunks for a given query, when different embedding models are used. As a first step, for a given dataset, we generate embeddings of queries and document chunks with each of the embedding models.  We then retrieve the $k$ most similar embeddings in terms of the cosine similarity for a particular query. As these embeddings correspond to specific chunks of text, we derive the sets of retrieved chunks C and C' for a pair of models. To measure the similarity of these sets, we use the Jaccard similarity coefficient as follows:

\begin{equation}
    Jaccard(C, C') = \frac{|C \cap C'|}{|C \cup C'|}
\end{equation}

Here, $|C \cap C'|$ corresponds to the overlap in text chunks by counting how often the two models retrieved the same chunks. Similarly, we can compute the union $|C \cup C'|$, which corresponds to all retrieved text chunks, counting chunks present in both sets only once. The resulting score is bounded in the interval [0, 1] with 1 indicating that both models retrieved the same set of text chunks.

While Jaccard similarity computes the percentage to which two sets overlap, it ignores the order in the sets. Rank similarity \cite{rank}, on the other hand, considers the order of common elements, with closer elements having a higher impact on the score. The measure assigns ranks to common text chunks according to their similarity to the query, i.e. $r_C(j) = n$ if chunk $j$ was the top-$n$ retrieved result for the query. Ranks are then compared using:

\begin{equation}
    Rank(r_C(j), r_{C'}(j)) = \frac{2}{(1 + |r_C(j) - r_{C'}(j)|)(r_C(j) + r_{C'}(j))}
\end{equation}

With this, rank similarity for two sets of retrieved text chunks C, C' is calculated as:

\begin{equation}
    RankSim(C, C') = \frac{1}{H(|C \cap C'|)}\sum_{j \in |C \cap C'|}Rank(r_C(j), r_{C'}(j))
\end{equation}

with $H(|C \cap C'|) = \sum_{k=1}^{K=|C \cap C'|}\frac{1}{k}$ denoting the K-th harmonic number, normalizing the score. Like the other measures, rank similarity is bounded in the interval [0, 1] with 1 indicating that all ranks are identical.

\section{Experimental Setup}

The following paragraphs describe our choice of datasets and models, along with details of the implementation of our experiments.

As we focus on the retrieval component of RAG systems, we select five publicly available datasets from the BEIR benchmark \cite{beir}. As generating embeddings for large datasets is a time-intensive process, especially for a larger number of models, we opt for five of the smaller datasets from the benchmark. This approach allows us to compare embeddings generated by a variety of models while at the same time allowing us to evaluate embedding similarity accross datasets. An overview of the datasets is shown in Table \ref{tab:datasets}. For each dataset, we create embeddings by splitting documents into text chunks such that each chunk contains 256 tokens. The embedding vectors are stored with Chroma DB \cite{chromadb}, an open source embedding database. For each vector, we additionally store information about the document and text chunk ids it encodes to be able to match embeddings generated by different models for evaluation.

For model selection, we primarily use publicly available models from the MTEB leaderboard \cite{mteb}. We do not simply pick the best performing models on the leaderboard; instead, our choices are influenced by several factors. Firstly, we focus on analyzing similarities within and across model families and pick models belonging to the e5 \cite{e5}, t5 \cite{sentencet5, gtrt5}, bge \cite{bge}, and gte \cite{gte} families. Secondly, we recognize that it might be of interest to users to avoid pay-by-token policies of proprietary models by identifying similar open-source alternatives. Therefore, we pick high-performing proprietary models, two from OpenAI (text-embedding-3-large and -small) \cite{openai} and one from Cohere (Cohere embed-english-v3.0) \cite{cohere}. We also compare the mxbai-embed-large-v1 (mxbai) \cite{mxbai} and UAE-Large-V1 (UAE) \cite{mxbaiuae} models, that not only report very similar performances on MTEB, but also identical embedding dimensions, model size and memory usage. Finally, we include SFR-Embedding-Mistral (Mistral) \cite{mistral} as the best-performing model on the leaderboard at the time of our experiments. A detailed overview of all selected models can be seen in Table \ref{tab:models}.

\begin{figure}[]
    \centering
    \includegraphics[width=\linewidth]{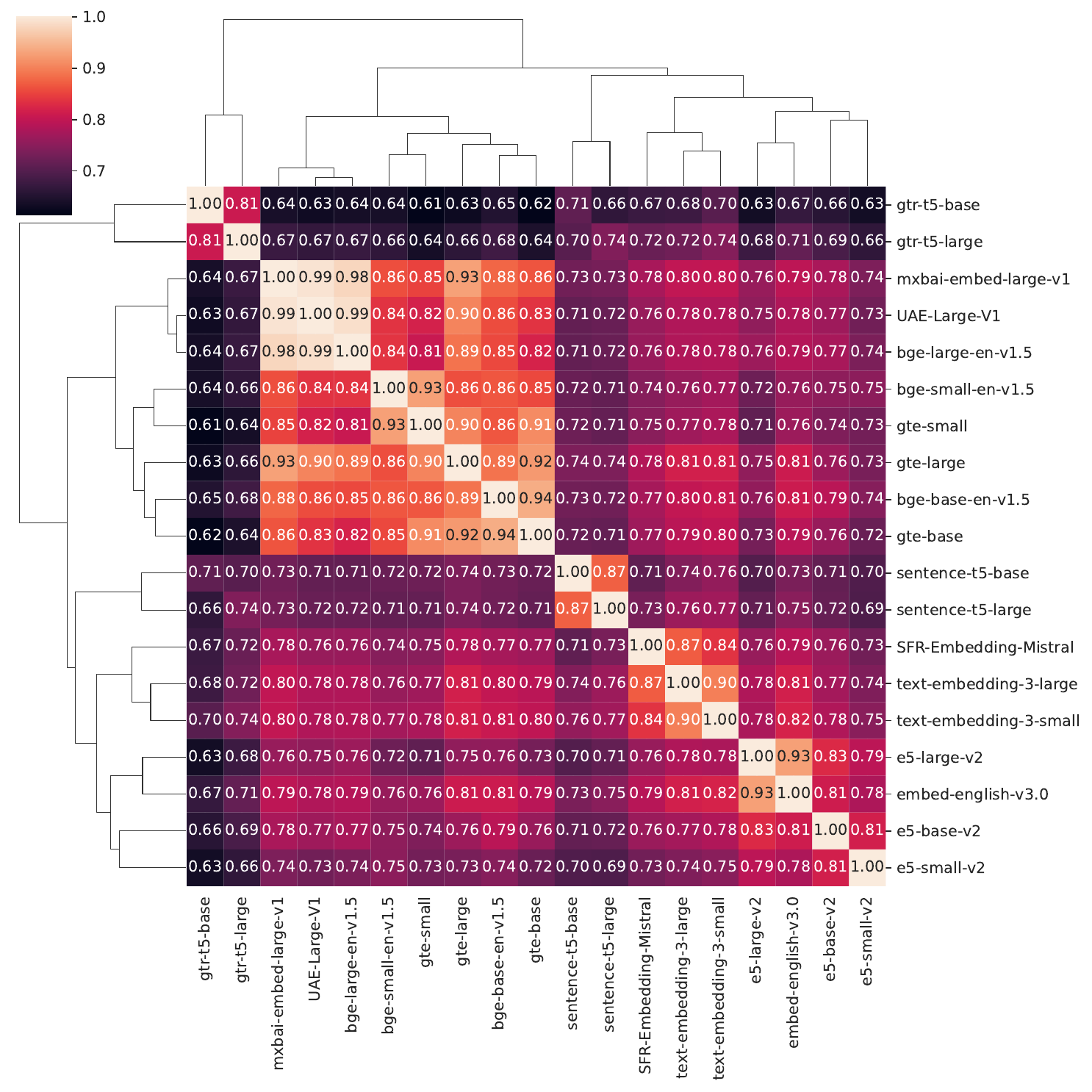}
    \caption{Mean CKA similarity across all five datasets. Models tend to be most similar to models belonging to their own family, though some interesting inter-family patterns are visible as well.}
    \label{fig:cka}
\end{figure}

\begin{figure}[]
    \centering
    \includegraphics[width=\linewidth]{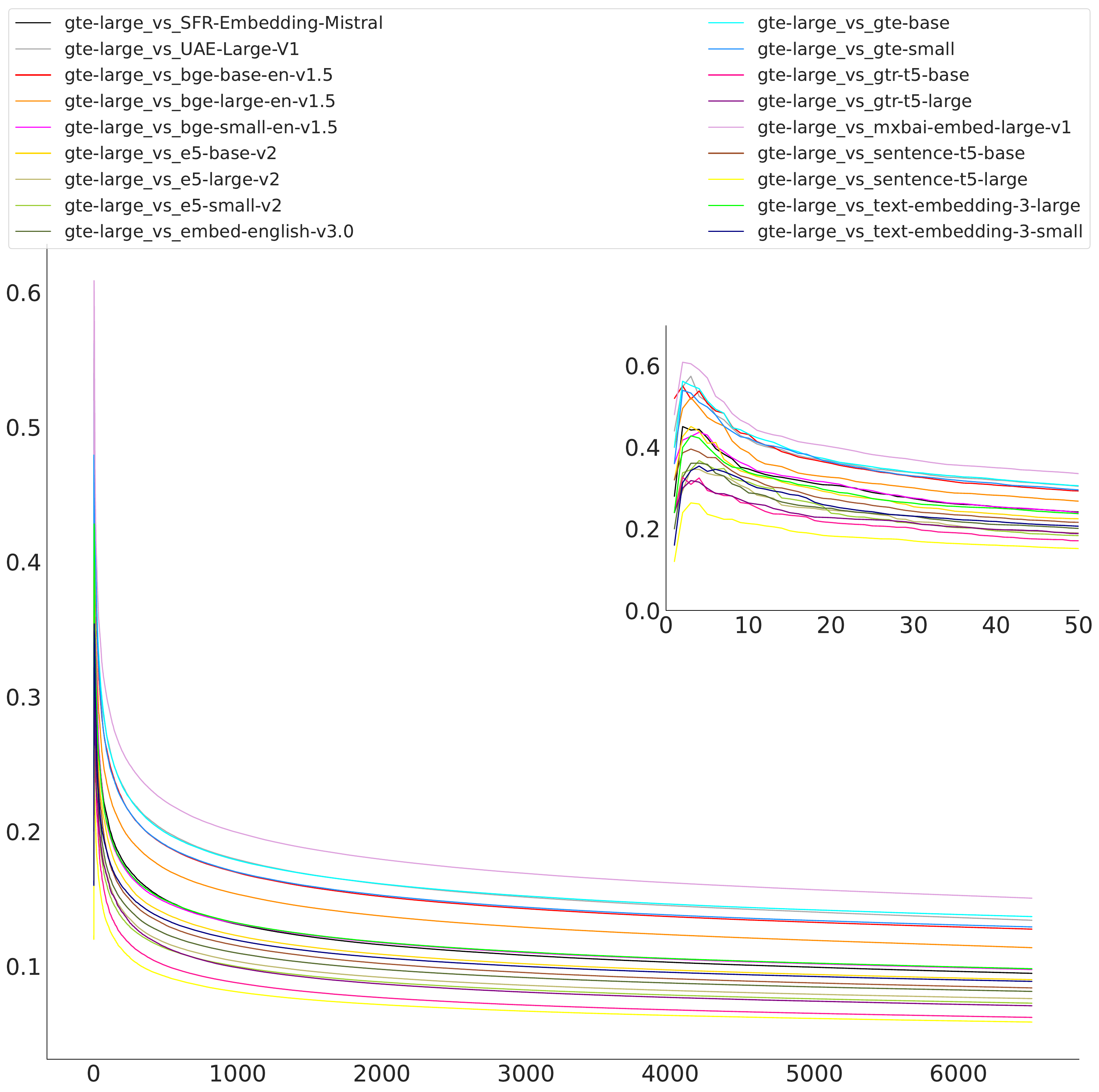}
    \caption{Rank similarity over all $k$ on NFCorpus, comparing gte-large to all other models. Scores are highest and vary most for small $k$, but then drop quickly before stabilizing for larger $k$.}
    \label{fig:rank_nf}
\end{figure}

To compare embedding similarity across models and datasets, we employ different strategies depending on the similarity measure. We apply CKA by retrieving all embeddings created by a model, matching embeddings using their document and text chunk ids and then computing their similarity for each of the five datasets. For Jaccard and rank similarity, we use sklearn's NearestNeighbor class \cite{scikit} to determine the the top-$k$ retrieval results. We compute Jaccard and rank scores per dataset, averaging over 25 queries. For the NFCorpus dataset, we calculate retrieval similarity for all possible $k$, i.e. using all embeddings generated for the dataset. As calculating similarity for each possible $k$ is computationally expensive, we did not repeat this for the remaining datasets and chose a smaller $k$ value instead. Furthermore, as only a limited number of results are to be provided as context to the generative model, analyzing retrieval similarity at low $k$ values for e.g. top-10 is of most interest. As we are interested in identifying clusters of similar models, we also perform a hierarchical clustering on heatmap values using Seaborn \cite{seaborn}. The following section describes the results of our evaluation for the different measures.

\section{Results}

To evaluate how similar embeddings generated by different models are, we will first consider model families, checking if their pairwise and top-k similarity scores are highest within their family. Subsequently, we will identify the open source models which are most similar to our chosen proprietary models.

\subsection{Intra- and Inter-Family Clusters}

\begin{figure*}
    \centering
    \subfigure[]{\includegraphics[width=0.48\textwidth]{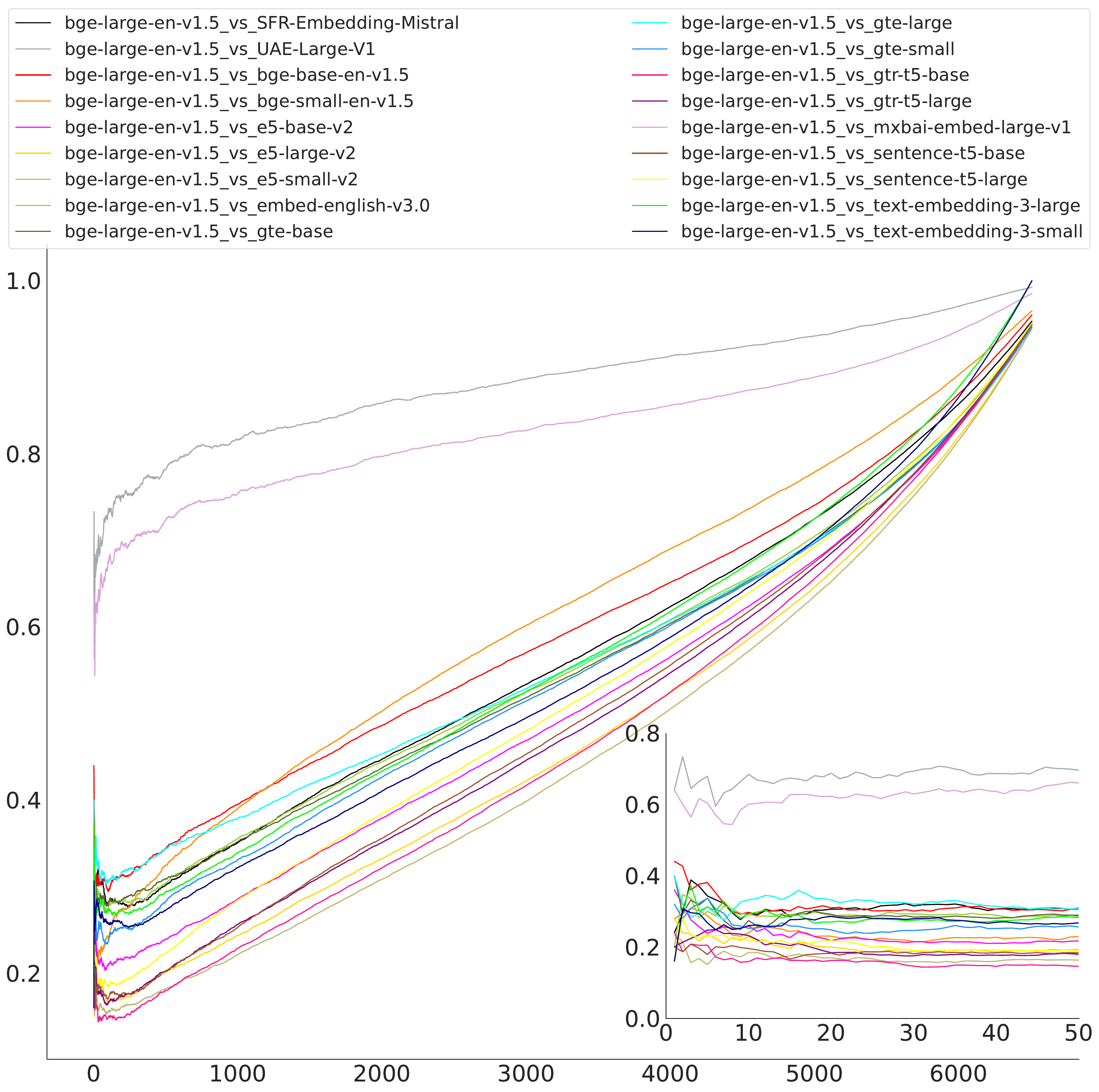}}
    \subfigure[]{\includegraphics[width=0.48\textwidth]{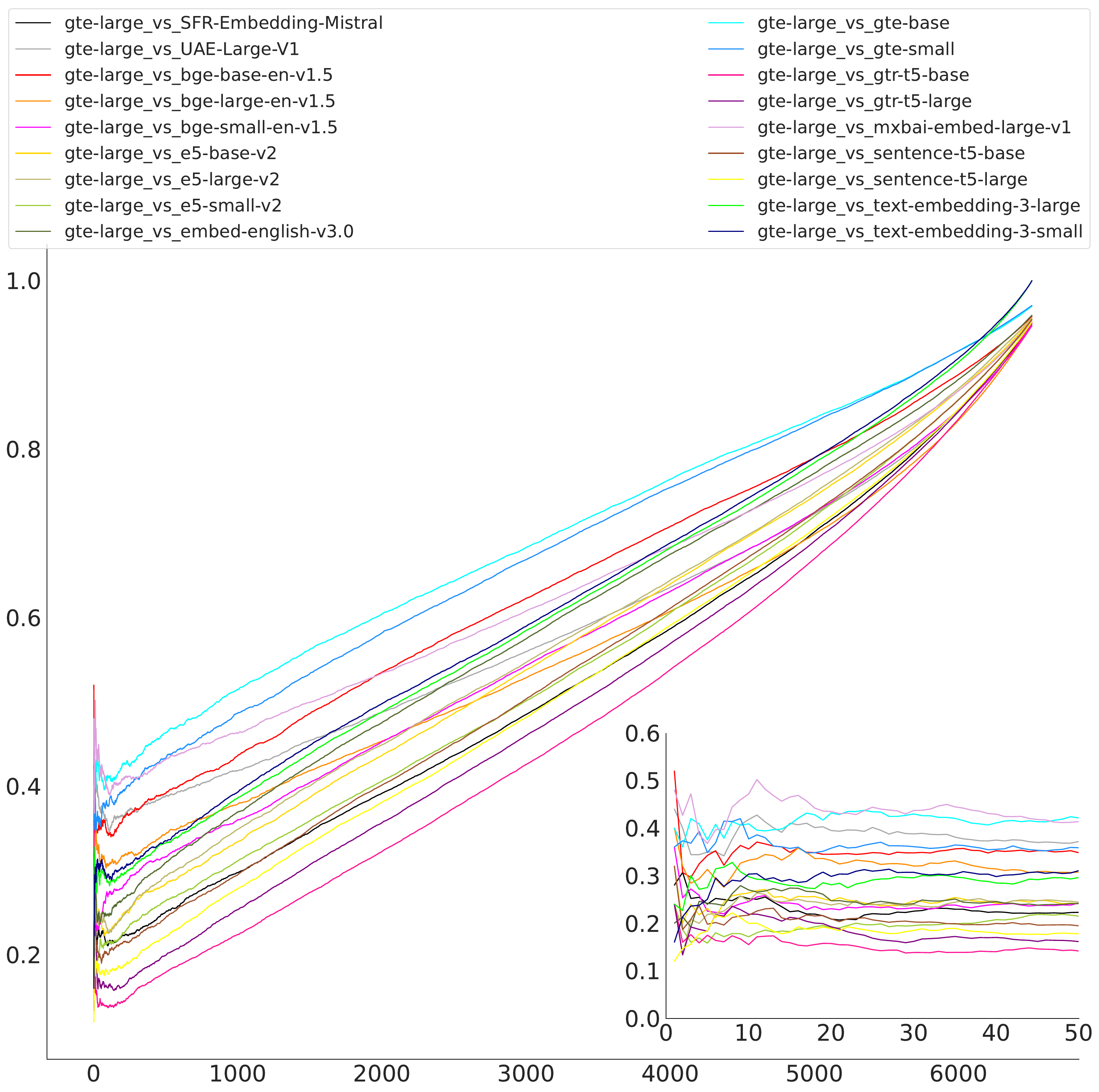}}
    \caption{Jaccard similarity over all $k$ on NFCorpus, comparing bge-large (a) and gte-large (b) to all other models. While bge-large shows high similarity to UAE-Large-v1 and mxbai-embed-large-v1, scores for gte-large are clustered much closer. Jaccard similarity seems to be most unstable for small values of $k$, which would commonly be chosen for retrieval tasks.}
    \label{fig:jac_nf}
\end{figure*}

Comparing embeddings directly with CKA shows high similarity across most of the models, albeit with some variance. These scores allow us to identify certain clusters of models. Figure \ref{fig:cka} shows the pair-wise CKA scores of all models averaged across the five datasets. As expected, scores for most models are highest within their own family. This holds true for the gtr-t5, sentence-t5 and text-embedding-3 (OpenAI) models. Although the sentence-t5 and gtr-t5 models are closely related, they do not exhibit significantly higher similarity with each other compared to the remaining models.

From an inter-family perspective, we observe high similarity between the bge and gte models. For some models in these two families, interestingly, the highest similarity scores rather correspond to inter-family counterparts with matching embedding dimensions than with models in the same family. Specifically, gte-small reports the highest similarity to bge-small and gte-base to bge-base. On the other hand, gte-large shows slightly higher similarity to bge-base than bge-large and thus to a model with a lower embedding dimension. Another inter-family cluster is formed by the three models with the highest CKA scores overall, namely UAE, mxbai and bge-large, whose scores suggest almost perfect embedding similarity. In fact, the similarity score of bge-large to these two models is much higher than to other bge models.

\begin{figure*}
    \centering
    \subfigure[]{\includegraphics[width=0.48\textwidth]{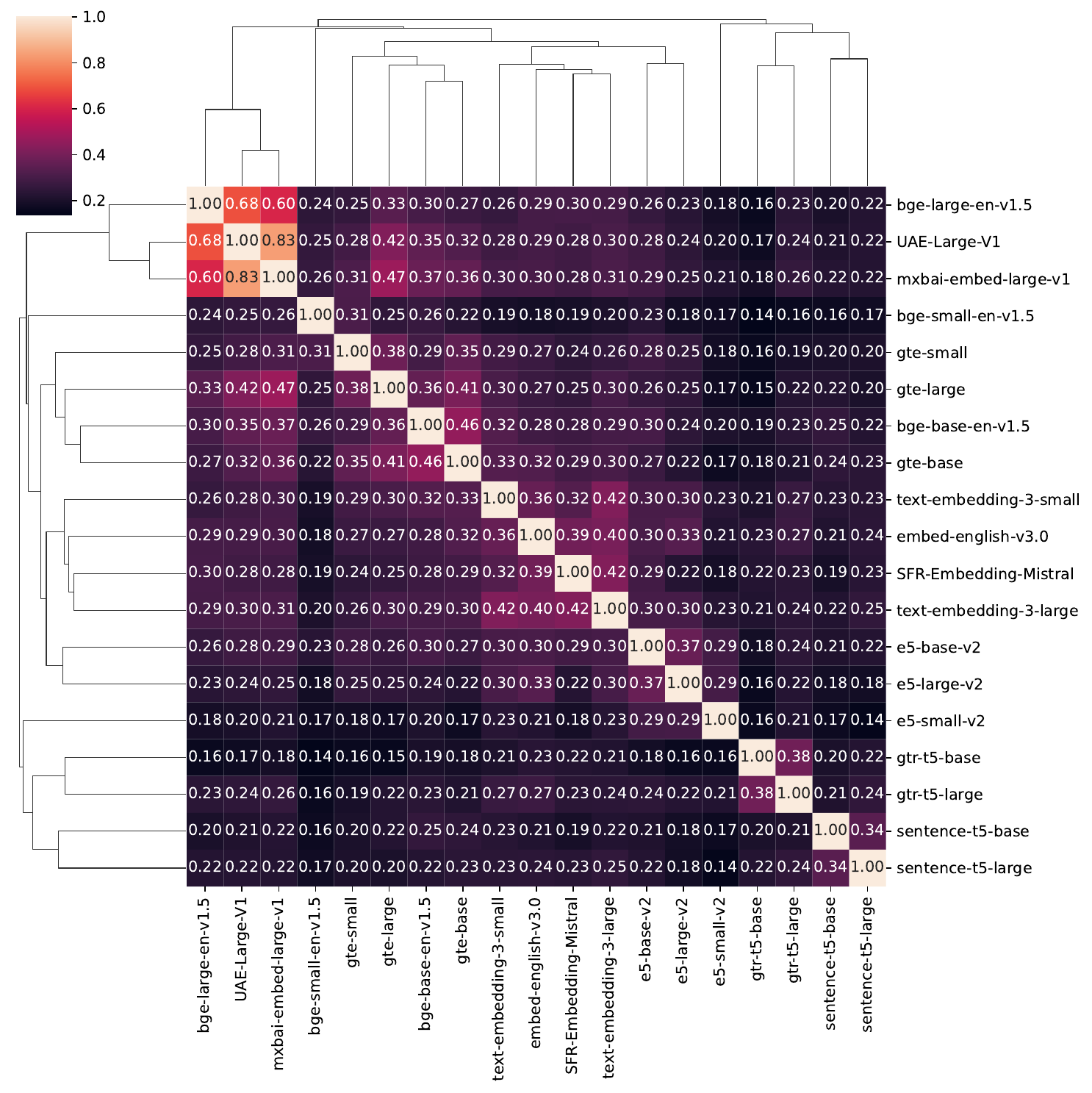}}
    \subfigure[]{\includegraphics[width=0.48\textwidth]{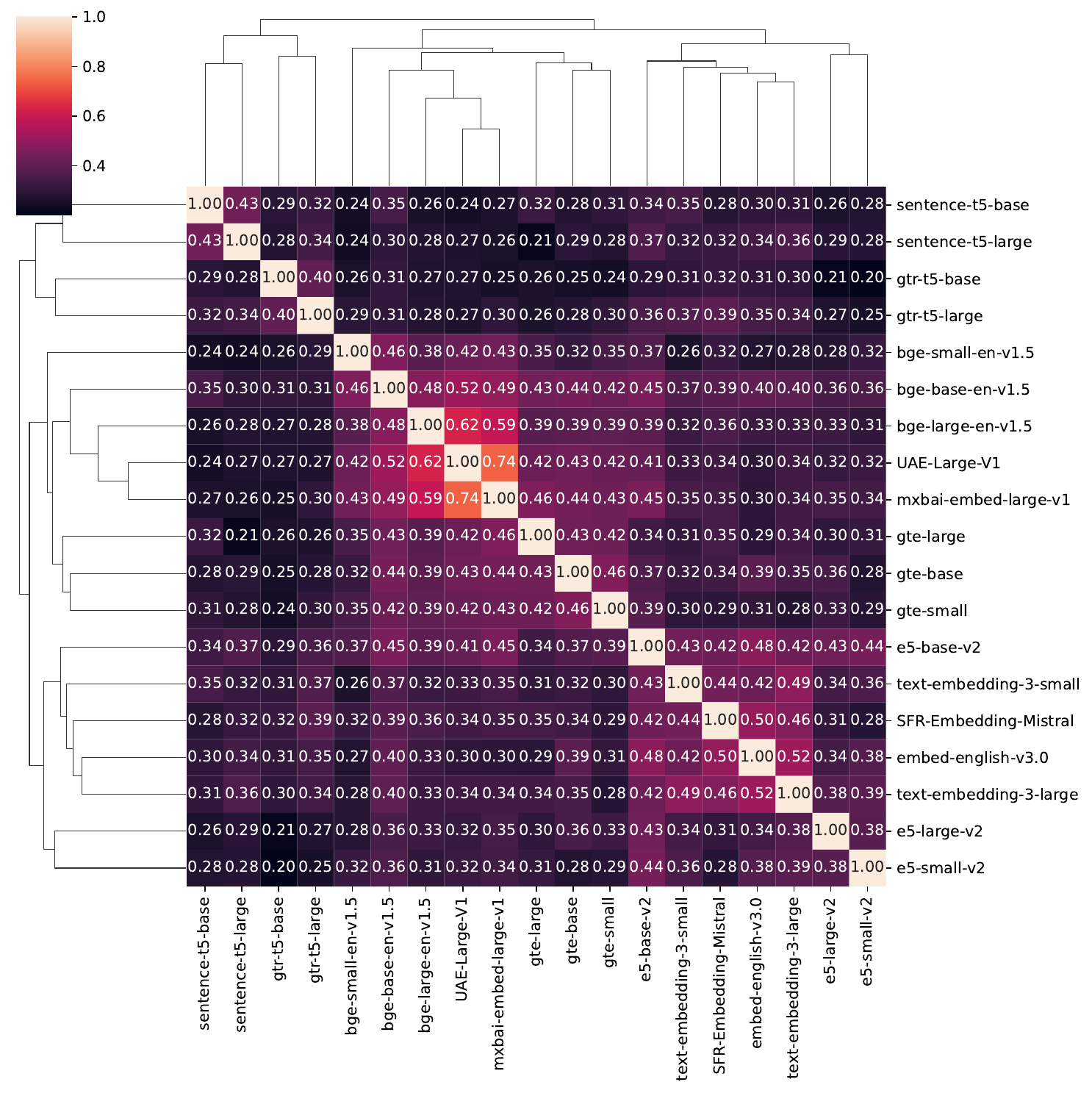}}
    \caption{Jaccard (a) and rank similarity (b) for the top-10 retrieved text chunks averaged over 25 queries on NFCorpus. The clusters vary slightly depending on the measure, as do the scores. Models tend to be most similar to models from their own family. However, some inter-family clusters are visible as well.}
    \label{fig:nf_10}
\end{figure*}

Shifting our attention to top-$k$ retrieval similarity, clusters vary depending on the $k$ value. Figure \ref{fig:jac_nf} illustrates how Jaccard similarity evolves over $k$ on NFCorpus. The first plot displays Jaccard scores between bge-large and all other models, while the second plot illustrates the scores for gte-large. For extremely low $k$, we observe some peaks for nearly all models, followed by a noticeable drop in similarity. Of course, for larger $k$, the scores converge to one. Re-affirming our earlier observations with the CKA metric, bge-large demonstrates high retrieval similarity with UAE and mxbai. Similarity to the remaining models is much lower, with the highest scores for bge-base and bge-small for larger $k$. However, especially for small $k$, there is high variance in similarity score, with models from other families, e.g. Mistral or gte-large sometimes achieving higher scores than the bge models. A similar pattern can also be observed in the second plot, where Jaccard similarity for gte-large is highest within its family for larger $k$, but models like mxbai or bge-base sometimes reporting higher similarity for small $k$. Therefore, the clusters we identified through our CKA analysis are only truly reflected in these plots for large values of $k$. This suggest that in real-world use cases, where the top-$k$ are crucial, such representational similarity measures might not provide the full picture. The plots for other model families provide nearly identical insights as those in the second plot in Figure \ref{fig:jac_nf} and thus we do not present them for sake of brevity. 

For rank similarity, scores peak for small $k$ and then quickly start to drop until they approach a low stable score for larger $k$ as shown in Figure \ref{fig:rank_nf} for gte-large. Once again, the bge/UAE/mxbai inter-family cluster shows the highest similarity. In contrast to Jaccard similarity, the clusters that could be observed for CKA do not always show for rank similarity. As can be seen in Figure \ref{fig:rank_nf}, the model with the highest rank similarity to gte-large is mxbai, rather than another gte model. Even so, the previously observed clusters also tend to appear for rank similarity, though they vary more depending on the models and dataset. Generally, scores for nearly all models are rather small for larger $k$, indicating low rank similarity. For small $k$, results vary more and differences between individual models are more pronounced.

\begin{figure*}
    \centering
    \subfigure[]{\includegraphics[width=0.48\textwidth]{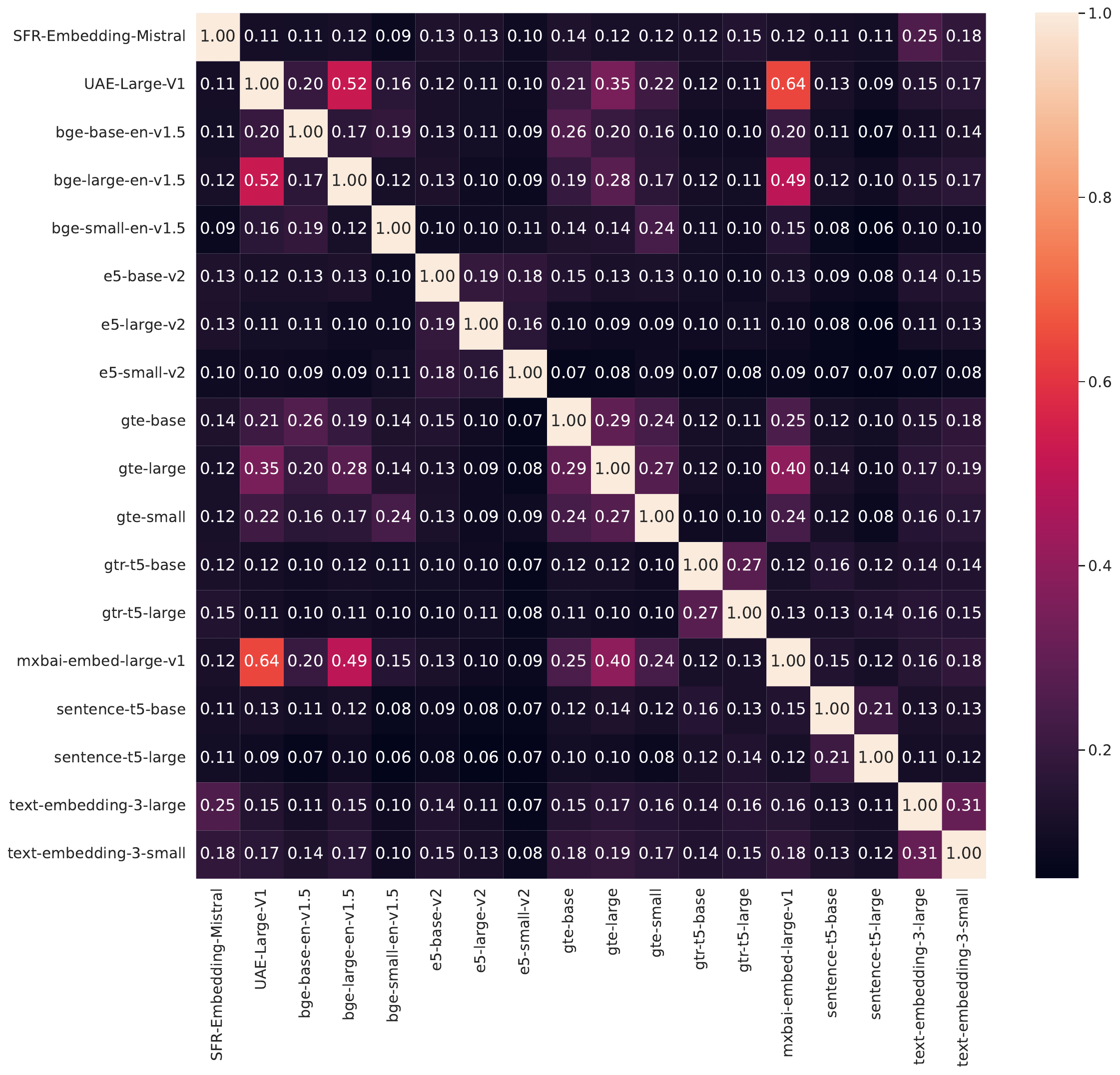}}
    \subfigure[]{\includegraphics[width=0.48\textwidth]{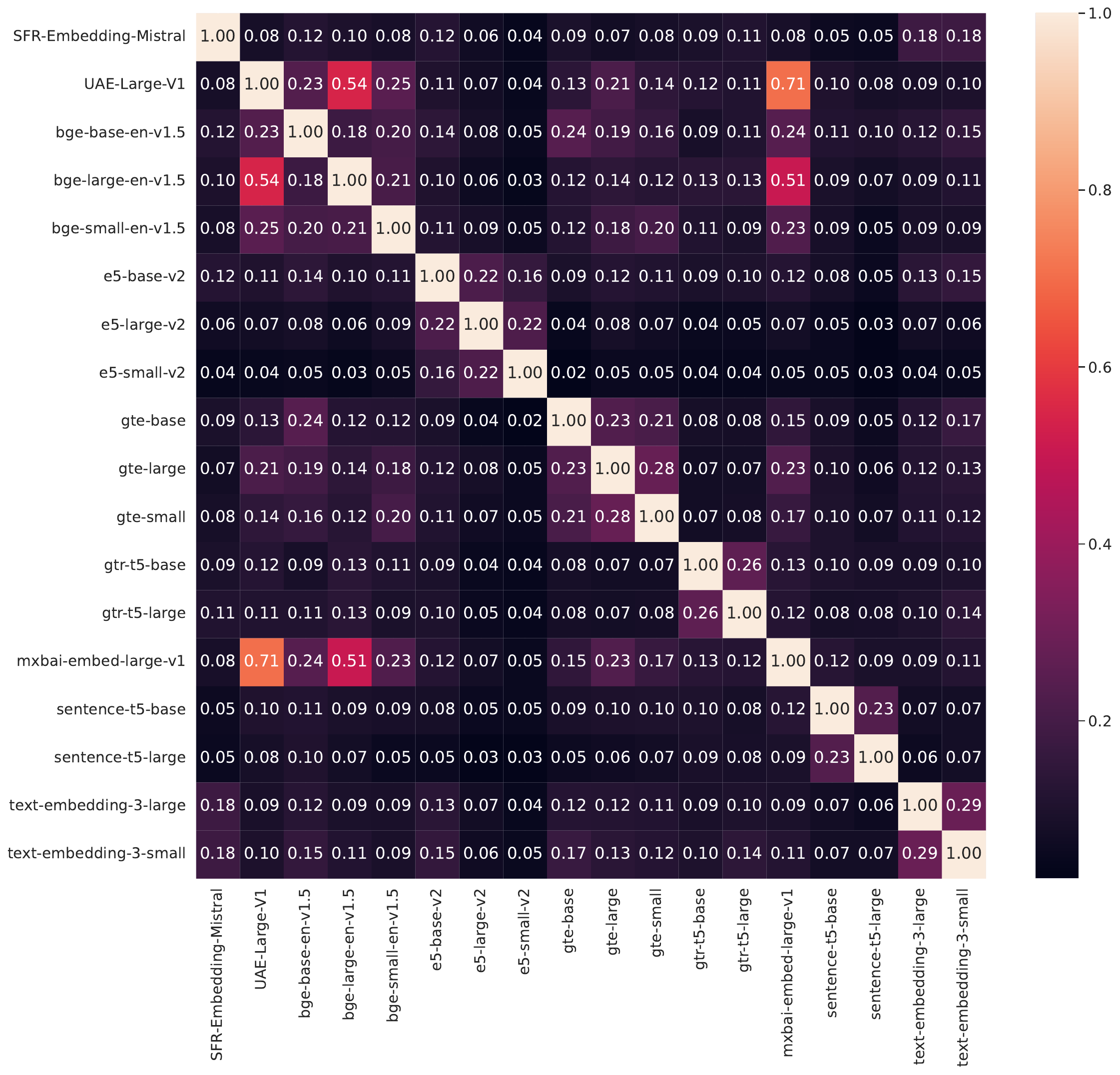}}
    \caption{Jaccard similarity for the top-10 retrieved text chunks averaged over 25 queries on FiQA-2018 (a) and TREC-COVID (b). Most models seem to retrieve almost completely distinct text chunks. Only the bge/UAE/mxbai cluster still shows a notable level of similarity, while the scores of the remaining clusters indicate only moderate to low levels of similarity.}
    \label{fig:fiq_cov_10}
\end{figure*}

As retrieval similarity at small $k$ is of most interest from a practical perspective, we take a closer look at top-10 Jaccard similarity. The heatmaps in Figures \ref{fig:nf_10}-\ref{fig:fiq_cov_10} show the top-10 Jaccard similarity between models across datasets. A striking insight here is that even the most similar models only report a Jaccard similarity of higher than 0.6, with the majority less than 0.5. This is of great relevance to practitioners, as it would imply that even using embeddings from models that report high representational similarity scores may yield little overlap in retrieved text chunks. As earlier, the cluster of UAE/mxbai/bge-large is the most prominent one with the highest scores. Intra-family scores tend to be the highest for most models, i.e. t5 and OpenAI. Depending on the dataset, this also applies to gte and e5 models, although Jaccard similarity to models from other families is sometimes higher. We also note that for the two larger datasets FiQA-2018 and TREC-COVID, the similarity scores are generally substantially lower, as can be seen in Figure \ref{fig:fiq_cov_10}. For the smaller datasets, Jaccard similarity is generally higher, though results differ depending on the data (see Figures \ref{fig:nf_10} and \ref{fig:sci_arg_10}). Similar observations can be made for rank similarity, although the appearance of family clusters is more dependent on the dataset. Larger datasets also lead to lower scores. These results illustrate that while family clusters can still be perceived at small $k$, they are not as prominent as they are for larger $k$. Furthermore, the top-10 retrieved results differ substantially for most models and datasets and their similarity might be dependent on the dataset itself.

\subsection{Open Source Alternatives to Proprietary Models}

We explicitly included proprietary models in our analysis to check which open source models are the best - which in our case means the most similar - alternative. The CKA scores in Figure \ref{fig:cka} indicate that embeddings generated by OpenAI's models (text-embedding-3-large/-small) are highly similar to those generated by Mistral, while the Cohere model (embed-english-v3.0) demonstrates high similarity to e5-large-v2.

\begin{figure*}
    \centering
    \subfigure[]{\includegraphics[width=0.48\textwidth]{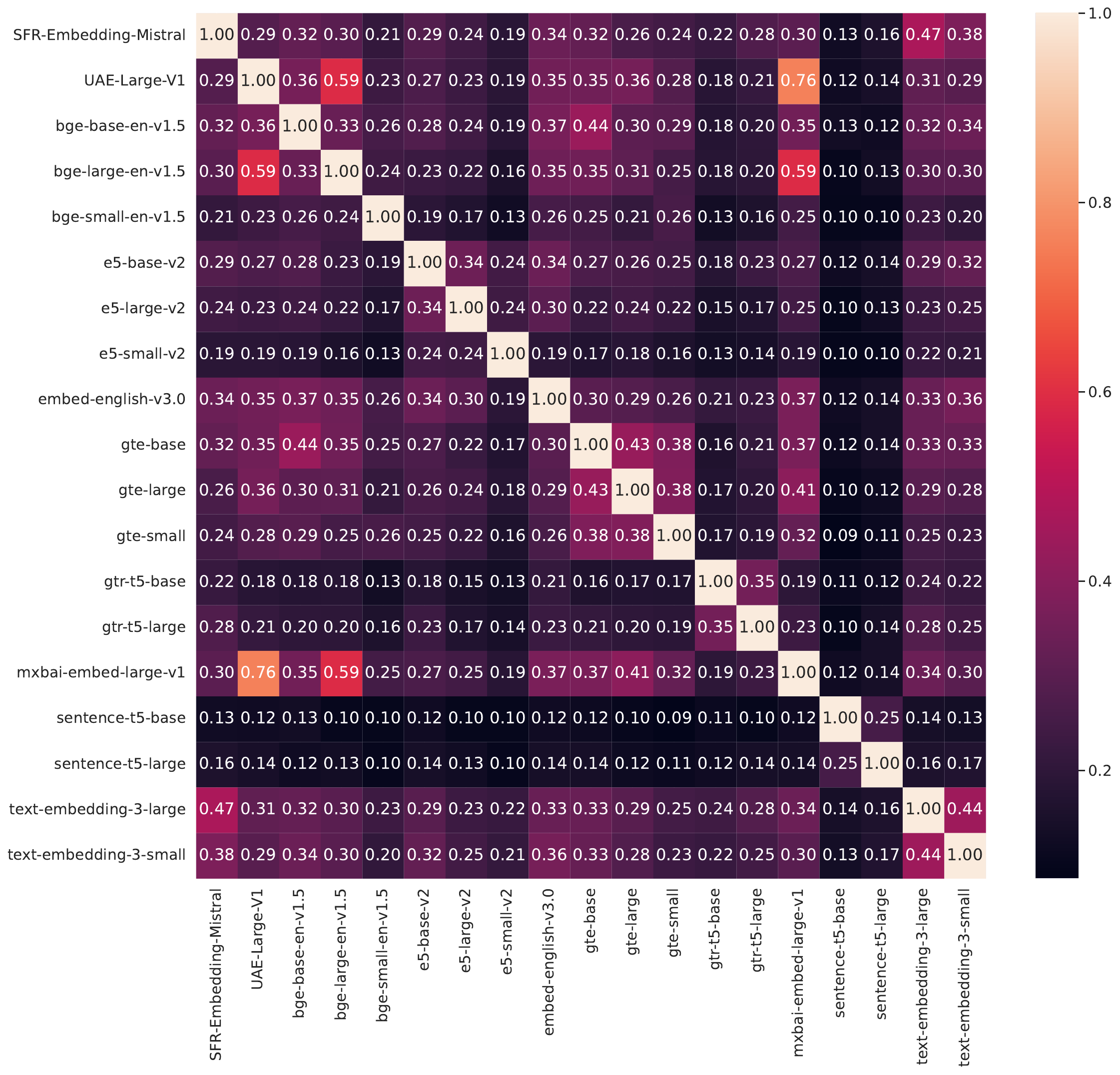}}
    \subfigure[]{\includegraphics[width=0.48\textwidth]{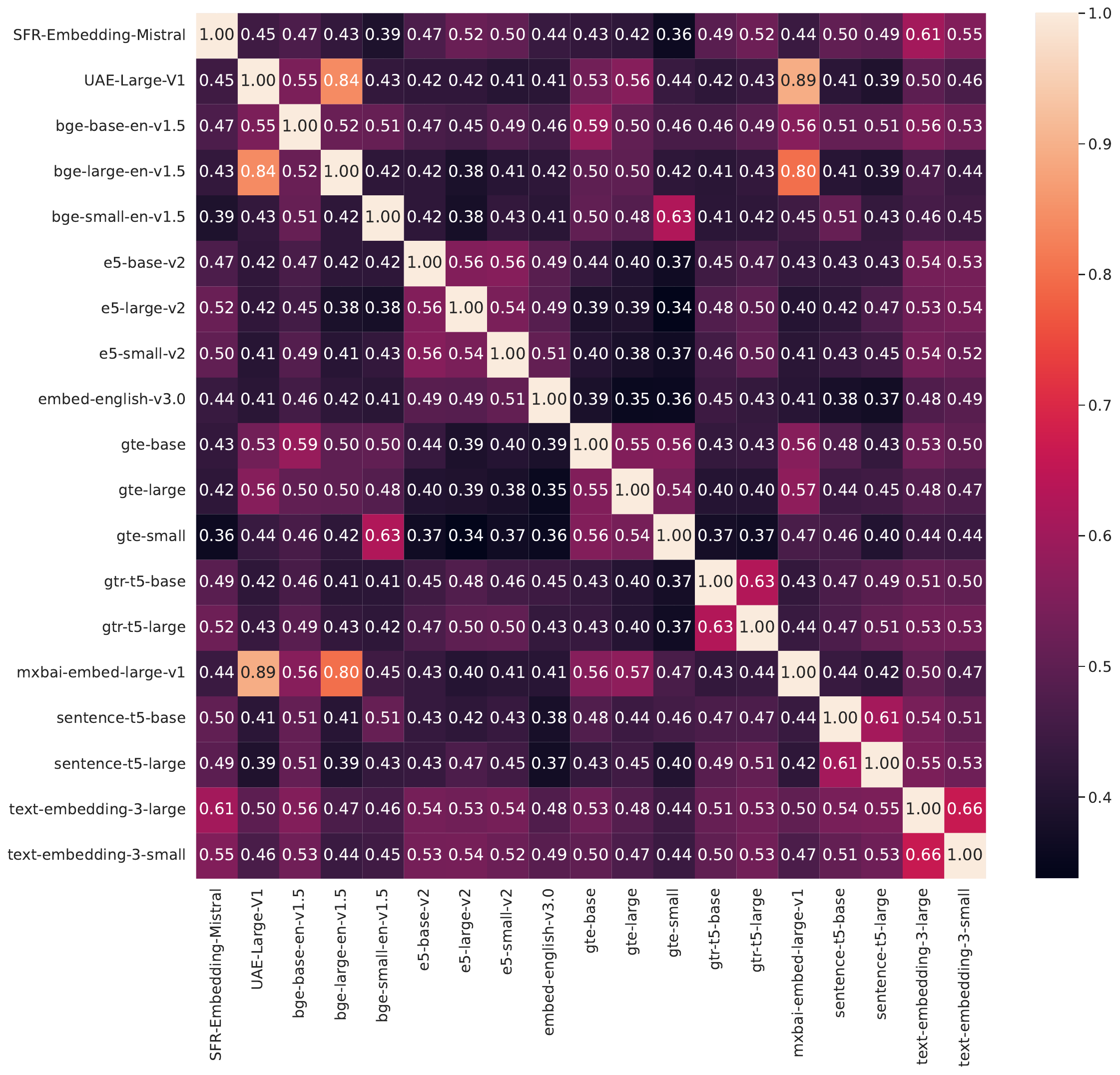}}
    \caption{Jaccard similarity for the top-10 retrieved text chunks averaged over 25 queries on SciFact (a) and ArguAna (b). The UAE and mxbai models show high levels of similarity along with bge-large. The remaining models tend to show the highest similarity within their own family with the exception of the bge/gte inter-family cluster.}
    \label{fig:sci_arg_10}
\end{figure*}

These observations do not entirely extend to retrieval similarity, especially for Cohere. While Mistral is still the most similar model to OpenAI's for larger $k$ across all datasets, there is no consistently most similar model for Cohere. Rather, the model varies depending on the dataset and measure - Jaccard and rank similarity - sometimes showing highest similarity to e5-large-v2, but sometimes also to other models like Mistral. Taking a closer look at top-10 similarity, Mistral still largely exhibits the highest similarity to the OpenAI models, especially to text-embedding-3-large. For text-embedding-3-small, scores on all datasets are rather close to those of other models. In absolute terms, however, retrieval similarity between Mistral and OpenAI models is only low to moderate. On smaller datasets, the highest Jaccard similarity to text-embedding-3-large only reaches about $0.6$ (see Figure \ref{fig:sci_arg_10}), while on TREC-COVID, the largest dataset, Jaccard similarity goes down to merely $0.18$ (see Figure \ref{fig:fiq_cov_10}). For Cohere's model, the most similar model for top-10 Jaccard similarity is different for each dataset, with the highest scores of $0.51$ occurring on ArguAna shwon in Figure \ref{fig:sci_arg_10}. For all proprietary models, even the best retrieval similarity at top-10 still suggests that the embeddings that would be presented to an LLM can differ notably. Once again, we could also observe dataset-dependent variance in scores, with lower retrieval similarity on larger datasets.

\section{Discussion}

While a pair-wise comparison of embeddings using CKA shows intra- and inter-family model clusters, retrieval similarity over different $k$ offers a more nuanced picture. Especially for small $k$, which are of most interest from a practical perspective, retrieval similarity varies. When comparing the top-10 retrieved text chunks, the low Jaccard similarity scores indicate little overlap in retrieved chunks, even when CKA scores are high. Especially for the two larger datasets FiQA-2018 and TREC-COVID, these scores are extremely low. As RAG systems usually operate on millions of embeddings, our datasets are comparatively small. Therefore, should a general trend of larger datasets leading to lower retrieval similarity exist, text chunks retrieved by different models in a regular use case might be nearly distinct for small $k$. Overall, our results suggest that even though embeddings seem rather similar when compared directly, retrieval performance can still vary substantially, is most unstable for $k$ values that are commonly used in RAG systems and also dataset-dependent. Retrieved text chunks at small $k$ show the least overlap, often leading to high differences in the data that would be presented to an LLM as additional context.

Our analysis demonstrates that although models tend to be most similar to models from their own family, inter-family clusters exist. The most prominent of these clusters is formed by the models bge-large-en-v1.5, UAE-Large-V1 and mxbai-embed-large-v1, which demonstrate high similarity even for retrieval at low $k$. Nevertheless, the high variance of retrieval similarity of the remaining clusters for small $k$ means that while the identified clusters might provide some measure of orientation when choosing an embedding model, the choice still remains a non-trivial task. Identifying suitable alternatives to proprietary models is likewise not as simple. While we were able to determine SFR-Embedding-Mistral as the model being most similar to OpenAI's embedding models, Jaccard similarity at top-10 for larger datasets shows a low overlap in retrieved text chunks. Furthermore, for Cohere's embedding model, we were unable to find a single most similar model, as this model varied across datasets for small $k$ values.

\section{Conclusion}

In this paper we evaluated the similarity of embedding models on different datasets. Given the large number of available models, identifying clusters or families of models with similar embeddings can simplify the model selection process. While previous work on LLM similarity exists, to the best of the authors' knowledge, it so far lacks a clear focus on embedding models specifically in the context of RAG. We therefore analyzed the similarity of embeddings generated by 19 different models using CKA for pairwise comparison as well as Jaccard and rank similarity to compare retrieval behavior at top-$k$ across five datasets. Comparing embeddings with CKA generally showed intra- and inter-family clusters across datasets. These clusters also appeared when evaluating top-$k$ retrieval similarity with large $k$ values. However, scores for low $k$ values, which would commonly be chosen in RAG systems, show high variance and much lower similarity, especially on larger datasets. Although we were able to identify some model clusters, our results suggest that choosing the optimal model remains a non-trivial task that requires careful consideration.

Still, we argue that a better understanding of how similarly different embedding models behave is an important research topic that requires further attention. There are a plethora of real-world scenarios where RAG systems can potentially be deployed. One such scenario, for example, is to retrieve relevant web content in response to a query. As large corpora of such data are available in the form of Web ARChive (WARC) files, evaluating embedding model similarity on such large, uncleaned datasets would perhaps lead to a better estimation of model similarity for a realistic RAG use case. Additionally, as documents often need to be chunked into smaller parts to fit into the models, the effect of chunking strategies such as token-based or semantic chunking on embedding similarity could be explored. Furthermore, our evaluation focused on a small sample of similarity measures, with their own definition about which criteria make models similar. Introducing more measures with different perspectives could improve our understanding on which factors influence model similarity. Finally, our focus was on identifying clusters or families of models, which for our initial experiments led us to choosing families of embedding models with varying performance on MTEB. With the frequent appearance of new models on the leaderboard and the focus on good MTEB performance, it would be of interest to compare the best performing models on MTEB and check if their relative difference in performance correlates with how similar these models are.

\begin{acks}
This work has received funding from the European Union's Horizon Europe research and innovation program under grant agreement No 101070014 (OpenWebSearch.EU, \url{https://doi.org/10.3030/101070014}).
\end{acks}

\bibliographystyle{ACM-Reference-Format}
\bibliography{references}

\end{document}